\documentclass[12pt]{article}
\usepackage{epsf}
\usepackage{cite}
\newcommand{\beqn}{\begin{eqnarray}}
\newcommand{\eeqn}{\end{eqnarray}}
\newcommand{\beq}{\begin{equation}}
\newcommand{\eeq}{\end{equation}}
\newcommand{\bdis}{\begin{displaymath}}
\newcommand{\edis}{\end{displaymath}}

\def\m{M_*}

\def\mpl{M_{\rm Pl}}

\newcommand{\gsim}{\lower.7ex\hbox{$
\;\stackrel{\textstyle>}{\sim}\;$}}
\newcommand{\lsim}{\lower.7ex\hbox{$
\;\stackrel{\textstyle<}{\sim}\;$}}

\begin{document}
\begin{flushright}
NYU-TH/01/11/15 \\
TPI-MIN-01/52 \\
UMN-TH-2033/01\\
\end{flushright}

\vspace{0.1in}
\begin{center}
\bigskip\bigskip
{\large \bf See-Saw Modification of Gravity}

\vspace{0.5in}

{Gia Dvali\,$^a$, Gregory Gabadadze\,$^b$, Xin-rui Hou\,$^b$, Emiliano
Sefusatti\,$^a$}
\vspace{0.1in}

{\baselineskip=14pt \it
$^a$Department of Physics, New York University,
New York, NY 10003\\[1mm]
$^b$Theoretical Physics Institute,
University of Minnesota, Minneapolis, MN 55455\\}
\vspace{0.2in}
\end{center}

\vspace{0.9cm}
\begin{center}
{\bf Abstract}
\end{center}

We discuss a model in which the fundamental
scale of gravity is restricted to $10^{-3}$ eV.
An observable modification of gravity
occurs simultaneously at the Hubble distance
and at around 0.1 mm. These predictions can be tested
both by the table-top experiments and by  cosmological measurements.
The model is formulated as a brane-world theory embedded in
a space with two or more infinite-volume extra dimensions.
Gravity on the brane reproduces the four-dimensional laws at observable
distances but turns to the high-dimensional behavior at larger scales.
To determine the crossover distance we
smooth out the singularities in the Green's functions by taking into
account softening of the graviton propagator due to the
high-dimensional operators that  are suppressed by
the fundamental scale.  We find that irrespective of
the precise nature of microscopic gravity the ultraviolet and
infrared scales of gravity-modification are rigidly correlated.
This fixes the fundamental scale of gravity at $10^{-3}$ eV.
The result persists for nonzero thickness branes.

\vspace{0.1in}

\newpage
%\providecommand{\LyX}{L\kern-.1667em\lower.25em\hbox{Y}\kern-.125emX\@}

%%%%%%%%%%%%%%%%%%%%%%%%%%%%%% User specified LaTeX commands.

\section{Introduction}

The law of gravitational interactions is
tested experimentaly at distances above $0.1$ millimeter
\cite {adel}. Hence, gravity could change its behavior
at a scale which is as low as:
\beq
\m^{\rm exp}~\sim~(0.1~{\rm mm})^{-1}~\sim~10^{-3}~{\rm eV}~.
\label{UV}
\eeq
A theory of gravitation which predicts
a deviation from the conventional law  above $\m$,
such that gravity is still negligible above this scale
compared to the gauge interactions,
cannot be ruled out at present.

Another important piese of data which is relevant for
gravity comes from recent astrophysical measurements.
The data indicate that the expansion of the
univesre might be accelerating \cite {cc}.
The existence of the cosmic acceleration raises yet another
interesting question on  possible infrared (IR) modification
of gravity. Indeed, an alternative view might be that
the observed acceleration of the universe
is not because of the vacuum energy, but
due to the modification of the laws of gravity at
the Hubble distance. Therefore, there might be another
scale in the theory of gravitation
\beq
H_0^{-1}~\sim~10^{29}~{\rm mm}~,
\label{hubble}
\eeq
where $H_0$ denotes the present-day value of the
Hubble parameter.

The aim of the present work is to discuss a model
which predicts the modification of gravity
at distances of
order $10^{29}$ mm and as a consequence of this, the modification
at $0.1$ mm  ensues. As we will see, these two scales are intimately
related. As a result, the model restricts
the value of the fundamental scale of gravity to $10^{-3}$ eV.

A brane-world model, in
which extra dimensions have an infinite-volume and
the 4D laws of gravity are reproduced on a brane
due to the induced Einstein-Hilbert term,
predicts the modification of gravity
at very large cosmological distances \cite {dgp}.
The 4D Einstein-Hilbert term ``shields''
a brane observer from strong bulk gravity \cite{dgkn1}.
Consequently, the observer detects the conventional
4D gravity all the way up to large cosmological distances.
At those scales the effects of infinite-volume
extra dimensions take over and
the behavior  of gravity changes.
In a 5D  model one finds the following crossover scale \cite {dgp}:
\beq
r_c^{(D=5)}~\sim ~{{\mpl^2}\over \m^3}~,
\label{rc5d}
\eeq
where $\mpl$ is the 4D Planck mass and $\m$ denotes the
fundamental gravitational scale of a 5D bulk theory.
For large distances, $r\gg r_c^{(D=5)}$,
gravity is five-dimensional, while it is
four-dimensional when $r\ll r_c^{(D=5)}$.
The laws of the 4D gravity  are
valid all the way down to the distances of order $1/\m$.
However, below  this length scale
the effective theory of gravity breaks down.
As a result, there is a  lower bound on the scale $\m$
which comes from the accelerator, astroparticle,
and cosmological data, that is $\m \gsim 10^{-3}~{\rm eV}$ \cite {dgkn1}.

Therefore, in a 5D theory one finds  the laws of 4D gravity only
in the interval:
\beq
{1\over \m}~\lsim~r~\lsim~{{\mpl^2}\over \m^3}~.
\label{int}
\eeq
This scenario leads to novel astrophysical and
cosmological consequences and predictions which were found  in
Refs. \cite{dg,cedric,ddg,dick,dick1,vilenkin,kofinas,ddgv,
ian,lue}.
In fact, it allows for the accelerated universe
with no cosmological constant which is consistent with
the data \cite {ddg}.
Some of these results will be discussed below.
A similar phenomenon for photons was studied
in Ref. \cite{dgs} (see also Ref. \cite {akhmedov}).
In these models the behavior of gravity and gauge fields are alike.

Note that in order for the large distance modification
of gravity to have relevance to present day cosmology and astrophysics
the crossover scale should be
at, or below,  the observable size of the universe,
$r_c^{(D=5)}~\lsim ~10^{29}~{\rm mm}$. This imposes the
following restriction $\m~\gsim~ 10~{\rm MeV}$.
Therefore, in this scenario, gravity will simultaneously be modified at
the Hubble distance and at $( 10 ~{\rm MeV})^{-1}\sim 20~{\rm Fermi}$.

In the present paper we will argue that the higher dimensional
($D\geq 6$) generalization \cite {dg} gives rise to a different
dependence of the crossover scale on $\mpl$ and $\m$.
In particular, for the modification of gravity
at around $10^{29}~{\rm mm}$
we get $\m \sim 10^{-3}{\rm eV}$, and {\it vise versa}.
The fundamental scale of gravity is bound to be $10^{-3}$ eV.
Thus, this model makes very restrictive
predictions on the modification of gravity which
are testable in table-top experiments and
simultaneously can be testable in astrophysical and
the cosmological measurements.

There are additional reasons why this
model can be interesting to study\footnote{For models with warped
geometries and modification of gravity at large scales
see Refs. \cite {Kogan,GRS}.}.
The scale of $10^{-3}$ eV
is very close to the scale of the vacuum energy density \cite {cc}.
Although, this similarity
might be a coincidence, nevertheless, it is rather intriguing to
argue, following Ref. \cite {sundrum},
that the scale of the vacuum  energy density in the Universe
might be related to
the ultraviolet cutoff of the gravitational theory.
Moreover, theories with infinite-volume
extra dimensions might shed new light on the
cosmological constant problem since the bulk SUSY can be preserved
in these models unbroken \cite {dgpcc,witten}, \cite{dg}
(for discussions of these models in string theory see Refs.
\cite{kir,lowe,zura}).

The generalization to the $D=(4+N)$-dimensional models
with $N\ge 2$  is not completely trivial, however.
Certain additional singularities emerge in this
case \cite {dg} and they should be dealt with some care.
The presence of these singularities
is related to the fact that the Green's function of the d'Alambertian
in the space transverse to a 3-brane blows up at the origin
for \( N\geq 2 \), while it is regular for \( N=1 \).
One of the purposes of the present paper is
to deal carefully with these singularities.

In what follows we will cast the singularities
into the following two classes.

\begin{itemize}
\item Type I: These are the singularities which are related to
the fact that the brane
is approximated by a delta-function type source. They are
present as in the \( N=1 \), as well as in \( N\geq 2 \) cases.

\item Type II: These are the singularities
which are present for the \( N\geq 2 \) case;
they are related to the fact that the
transverse Green's function of the d'Alambertian in \( N\geq 2 \)
dimensional space blows up at the origin.
\end{itemize}

Although these two types of singularities have a common origin,
nevertheless, it is convenient
to discuss them separately in order to compare
the $N=1$ and $N\ge 2$ models.

Let us start with the Type I singularities, i.e., those associated with
a vanishing brane width. Suppose the brane had a finite thickness
(we call it $\Delta$)\footnote{
Below in sections 3-6 we will be dealing with a brane
the transverse width of which is much smaller
than the fundamental distance scale in the bulk. Such an unusual brane
can exist even in 4D theory,
see discussions below and in section 3 of Ref. \cite {dgkn1}.}.
What would be the effect of this width on
a low energy observer in the 4D worldvolume theory? One
effect is the presence of a massive state (or states) which
correspond to the fluctuation of the brane transverse
width (the {}``breathing mode{}'').
The mass of this state is of the order of the inverse brane width.
This state, being localized and gravitationally interacting,
will certainly renormalize the graviton kinetic term on the
4D worldvolume, so that the generic value of the
coefficient in front of the 4D graviton induced
term is of the order $\Delta^{-1}$. Other than
that, the presence of such a heavy state is irrelevant for
low-energy 4D physics\footnote{It also renormalizes the brane tension.}.
Therefore, from the point of view of a low-energy observer this state can be
{}``integrated out{}''. This is equivalent of {}``integrating out{}'' the
brane width. Hence, in the low-energy theory the brane will look as a
delta-function source.
Since we expect that the physics is smooth and continuous,
the calculations
at low energies, i.e., at \( E\ll \Delta^{-1} \), in a theory with the
delta-function brane should produce the same result
(up to the ${\cal O}(E\Delta)$ corrections)
as calculations in a theory where the ultraviolet
resolution of the brane is manifest due to a nonzero brane width.
This is indeed what we will find below.

Let us now turn to the Type II singularities.
They will  be taken care of
separately from the singularities associated
with the delta-function brane.
In fact, as a first step we could keep the brane
as a delta-function source and try to remove the Type II
singularities. For this one could introduce the
{}``\( \epsilon   \) shift regularization{}'' as it was done in \cite{dg},
or one could use the rigid cutoff in the bulk as in
Ref. \cite{wagner}. However,
a more justified way (that automatically preserves reparametrisation
invariance) to deal with this problem is to introduce higher dimensional
HD operators as was proposed in Refs. \cite{kir} and \cite {dgkn1}.
From the physical standpoint, the
HD's are remnants of UV physics which is generically unknown.
Thus, it is natural
that they provide regularization of certain UV singularities in the low-energy
theory. This is the way which we will follow in the present paper.

The paper is organized as follow.
In section 2 we study  the induced terms
on a thick  brane. We will show that these terms in general are
nonlocal and have no factorizable form in terms
of the worldvolume and transverse coordinates.
These nonlocal expressions, however,
can be expanded in an infinite series of
local terms. The expansion parameter is the width of the brane.
The leading term  coincides with the 4D Einstein-Hilbert term.
In section 3 we setup a scalar field theory model which
mimics the properties of a theory with the
induced kinetic term on a thick brane. At this point we fine-tune
the brane tension and the 4D cosmological constant to make the
calculation analytically treatable. In section 4 we use this model to
calculate the two-point
Green's function. First we consider the delta-function brane but
introduce the HD operator
which smoothes out the Type II singularities in the expression for the
Green's function. We show that the behavior of the gravity
is qualitatively similar to that of the 5D theory: at short
distances the
theory behaves as four-dimensional and at large distances it approaches the
\( (4+N) \)-dimensional regime. The expression
for the crossover distance in this case is different from the
one found for the 5D theory.
In section 5 we consider the case when the brane width is kept fixed
and calculate the  corresponding Green's functions.
We show that the qualitative behavior described
above holds unchanged.
The limiting transition to  the delta-function
brane is smooth. There are no ultra-local
interactions \cite {DR} in the bulk
and no new physical scales \cite {kir} arise in these
calculations.
In section 6 we discuss the phenomenological constraints on
the crossover scale beyond which gravity can change its regime.
In section 7 we consider a brane with a nonzero tension and
give an estimate for the crossover scale in this case.
Conclusions are presented in section 8.

\section{Induced Terms on Smooth Branes}

In this section we study how the induced
terms arise on a smooth solitonic brane.
We will argue that if a nonzero
brane width is kept, the induced terms are nonlocal
and not factorizable with respect to worldvolume and transverse dimensions.
However, in the low-energy approximation they
can be expanded in a power series of local terms
with the leading term coinciding with the 4D
Einstein-Hilbert action.

Our consideration in this section is rather general and applies
to any brane-world model (with a non-conformal-invariant
worldvolume theory). Let us start with a $(4+N)$-dimensional action
which contains a graviton $G_{AB}$ and a scalar field $\Pi$
which makes the brane, and other fields $\Psi$ which do not
participate in the formation of the classical background\footnote{
All the derivations in this section can be generalized straightforwardly
for more complicated classical backgrounds.}:
\beq
S~=~\int~d^4x~d^Ny~\sqrt{G}~{\cal L}(G,~\Pi,~\Psi)~.
\label{act}
\eeq
Here ${\cal L}$ denotes the total Lagrangian density
which is a function  of the fields and their derivatives
which are suppressed in (\ref {act}).
The bulk graviton is  decomposed as follows:
\beq
G_{AB}(x_\mu,y_i),~ A,B=0,1,...,3+N;~~~\mu,\nu = 0,1,2,3;~~~i,j=4,5,..,3+N~.
\label{G}
\eeq
We study  the Lagrangian which yields
a classical solution in a form of a three-brane.
This solution depends on the coordinates $y_i$ but does not
depend on $x_\mu$.   Many examples are known
in various dimensions. Let us split the graviton and scalar
fields in their  classical parts and fluctuations:
\beqn
G_{AB}~=~G_{AB}^{\rm cl}(y)~+~H_{AB}(x,y)~, ~~~~
G_{\mu\nu}^{\rm cl}(y)~\equiv~A^2(y)~\eta_{\mu\nu}~,~~G_{\mu j}^{\rm cl}~=~0~,
\label{gcl} \\
\Pi(x_\mu,y_i)~=~\Pi^{\rm cl}(y)~+~\sigma (x_\mu)~f(y)~.
\label{picl}
\eeqn
Where in (\ref {gcl}) we use the parametrisation
of the $\{\mu\nu  \}$ component of the background metric
in terms of the function $A^2(y)$ which is typical for the branes
that  preserve the 4D Poincare invariance on the
worldvolume.

The scalar field $\Pi$ has modes which are localized on a brane.
The most obvious ones are the
Goldstone bosons associated with spontaneously broken
translation invariance in the $y_i$ directions.
These particles are massless  and are derivatively coupled
to matter fields on a brane. For these reasons
they are not relevant for our discussions below.
In addition, generically there are massive localized modes
on a brane.  For simplicity we consider below only a single
mode. As we mentioned  before, the latter
corresponds to the fluctuations of the transverse
width of the brane. Thus, its mass is proportional to the
inverse brane width. Let us discuss this mode in detail.
Since it  is localized on a brane
there should exist an effective four-dimensional Lagrangian
for it. To derive this Lagrangian let us start with the
action for $\Pi$:
\beq
\int~d^4x~d^Ny~\sqrt{G}~\left \{ {1\over 2}~G^{AB}~\partial_A\,\Pi~
\partial_B\,\Pi~ -~ V(\Pi) \right \}~.
\label{actpi}
\eeq
Here $V$ is a potential which is responsible for the existence
of the brane.
Let us now substitute (\ref {gcl}) and (\ref {picl}) into
(\ref {actpi}) and keep truck of quadratic fluctuations
(interactions will be discussed later). In this approximation
we find:
\beq
\int~d^4x~d^Ny~\sqrt{G^{\rm cl}(y)}~ \left \{
{f^2(y)\over A^2(y)}~
{\partial_\alpha \sigma(x)\partial^\alpha \sigma(x)\over 2}~-~
{\sigma^2\over 2}~[f^2~V^{\prime\prime}(\Pi^{\rm cl})-
\partial_if\partial^if ]
\right \}~.
\label{massterm}
\eeq
(The  prime denotes differentiation w.r.t. $\Pi$.
The 4D indices here and below are contracted by the
flat space metric $\eta_{\mu\nu}$ and those of the extra
coordinates by $G^{\rm cl}_{ij}$.)
The quadratic term in $\sigma$ is a mass term.
The value of the mass  depends on the exact expression for
$f$. For instance, by definition of a localized Goldstone
particle its profile $f$ is such that the mass term in
(\ref {massterm}) vanishes.
On the other hand, the profile $f$ for a breathing mode is different,
it satisfies the equation
\footnote{Note that functions $f$ for different modes
are orthogonal to each other, that is why we can treat them
separately.}:
\beq
f^2~V^{\prime\prime}(\Pi^{\rm cl})~ +~
{f\,\partial_i \left (\sqrt{G^{\rm cl}(y)}\,\partial^if \right )\over
\sqrt{G^{\rm cl}(y)}}
 ~=~{M^2~f^2~\over A^2}~.
\label{mass}
\eeq
Here the parameter $M$ is proportional to the inverse brane width:
\beq
M~\sim~ {1\over \Delta}~.
\eeq
Using (\ref {mass})
we obtain  the following expression for the
quadratic part of the low energy 4D action of a
breathing mode:
\beq
\left [\int~~\sqrt{G^{\rm cl}(y)}{f^2(y)\over A^2(y)}~d^Ny \right ]
\times \int~d^4x~
\left \{
{1\over 2}~\partial_\alpha \sigma(x)\partial^\alpha \sigma(x)~-~
{1\over 2}~\sigma^2  M^2
\right \}~.
\label{quadr}
\eeq
We see that the $y$ dependence can be integrated out.
This is because for large $y$ the localization function
for local defects is exponentially decreasing,
($ \sim {\rm exp}(-|y|/\Delta)$). This has to make the
$y$ integral in (\ref {quadr}) to converge for a breathing mode
 even though the integrals
$$\int~~\sqrt{G^{\rm cl}(y)}d^Ny ~~~~{\rm and}~~~
\int~~\sqrt{G^{\rm cl}(y)}/{A^2(y)}~d^Ny$$
might not be convergent
(the latter case corresponds to the brane models where
there is no localized graviton zero mode).

Let us now turn to the gravitational part of the action.
We will consider a general case when  the graviton
is not necessarily localized. If so, the bulk gravity action
cannot be integrated w.r.t. $y$ to obtain 4D kinetic term
for 4D gravitons. Nevertheless, the 4D laws of gravity can be obtained
on a brane due to the induced terms \cite {dgp,dg} to the discussion of which
we turn now.

Let us look at the interaction of bulk gravity with the
localized field $\sigma$.
We concentrate first on the $\{\mu\nu \}$ components
of the interactions since these are the ones
that determine the 4D induced terms relevant to us.
Using (\ref {gcl}),(\ref {picl}) and (\ref {actpi}) we derive
the interaction Lagrangian for the  $\{\mu\nu \}$  part:
\beqn
\int~d^4x~d^Ny~\sqrt{G^{\rm cl}(y)}
{f^2(y)\over A^4(y)}~ \eta^{\alpha\mu} \eta^{\beta\nu} H_{\alpha\beta} (x,y)
~\times
\nonumber \\
\left \{
~\partial_\mu \sigma(x)\partial_\nu \sigma(x)
-\eta_{\mu\nu} \left [ {1\over 2}~
\partial_\lambda \sigma(x)\partial^\lambda \sigma(x)~-~
{1\over 2}~\sigma^2  M^2
\right ] \right \}~.
\label{square}
\eeqn
The quantity
in second line in (\ref {square}) is nothing but
the energy momentum tensor for $\sigma$ with the  mass $M$.
From this we see that the interaction of a bulk graviton with
the localized mode can be rewritten in a purely 4D form.
Indeed,  rescaling the
field $\sigma \to \sigma / (\int d^Ny \sqrt{G^{\rm cl}(y)}~f^2(y)/
A^2(y))^{1/2}$ we get  the canonically normalized 4D
kinetic term for the field $\sigma $ in (\ref {quadr})
and its interaction with the bulk gravity,
according to Eq. (\ref {square}),
takes the following form:
\beqn
S_{\rm int}
~=~\int~d^4x ~h^{\mu\nu}(x)~ T_{\mu\nu}^\sigma(x)~,
\label{g}
\eeqn
where the field $h$ is defined as follows:
\beqn
h_{\mu\nu}(x)~=~{\int~d^Ny~ [ \sqrt{G^{\rm cl}(y)}~{f^2(y)}~
H_{\mu\nu}(x,y)/A^4(y)]   \over
\int~d^Ny~  [\sqrt{G^{\rm cl}(y)}~{f^2(y)}/A^2(y)] }~.
\label{gg}
\eeqn
Thus, the interaction Lagrangian of higher dimensional gravity with
a localized mode can be given a purely  4D form (\ref {g})
in spite of the fact that the bulk graviton
might not be localized on a brane (i.e., the bulk graviton
kinetic term might not be integrable w.r.t. $y$).
In what follows we will study loop  effects due to this interaction.

Before we turn to these issues  we would like to digress
for a moment and  make  two comments. So far we discussed only a single
localized field. Suppose now there are $n$ localized fields
$\sigma_k~,k=1,2...,n$,
with the localization functions $f_k$ which in general
can be different. Then, the interaction vertex of each of this state
with the bulk gravity can be presented in the form of Eq. (\ref {g}).
On the other hand, since $f$'s  are different and the expressions for $f$'s
determine $h_{\mu\nu}(x)$ in (\ref {gg}), then
each of these localized states will interact
with different ``effective 4D gravitons'' (\ref {gg}).
However, the effects that discriminate between these gravitons
are suppressed by the brane width
and are negligible as we will see below.
For this let us perform  the  expansion:
\beq
H_{\mu\nu}(x,y) ~=~H_{\mu\nu}(x, 0)~+~y^i\, \partial_i H_{\mu\nu}(x, 0)~+~
{1\over 2}~y^i\,y^j\,  \partial_i \,\partial_j H_{\mu\nu}(x, 0)+...~.
\label{exp}
\eeq
We substitute this series into (\ref {gg}) and then in (\ref {g})
for each localized field. As a result, for each field we
get infinite number of terms first two of which are:
\beqn
H^{\mu\nu}(x,0)~ T_{\mu\nu}^{\sigma_k}(x)~\int~\sqrt{G^{\rm cl}(y)}
~{f_k^2(y)\over A^2(y)}~d^Ny, \label{0} \\
\partial_iH^{\mu\nu}(x,0)~ T_{\mu\nu}^{\sigma_k}(x)~\int y^i~
\sqrt{G^{\rm cl}(y)}
~{f_k^2(y)\over A^2(y)}~d^Ny, \label{1}
\eeqn
and so on, with increasing powers of derivatives.
The first term (\ref {0}) gives the universal interaction for all
the different fields. This can be seen
by performing the rescaling of the fields to bring their
kinetic terms in (\ref {quadr}) to the canonical form.
Then, the couplings of all the
localized fields to $H_{\mu\nu}(x,0)$
are identical. In fact, the result of this universal
term is what we obtain
in the delta-function limit for the brane, i.e., when
$f^2\to \delta^{(N)}(y)$.  This is the dominant contribution.
The higher terms,  on the other hand, are not like this.
Each higher term (including the one in (\ref {1}))
can be thought of as a new field
$\partial_i\partial_j...\partial_k ~H_{\mu\nu}(x,0)$
from the 4D point of view.
However, as we can see from (\ref {1}),  the
couplings of these fields to the energy-momentum tensor
contain extra powers of $y$ under the integral.
Since the effective region of integration is determined by
the ``width'' of the function $f$, then each additional power of $y$
will translate into an additional power of $\Delta$.
Therefore
these couplings are suppressed by extra powers of the brane
width $\Delta$. For instance, the term in (\ref {1}) can give rise to
the contributions containing extrinsic curvature and its derivatives.
However, these terms are suppressed compared to the leading ones
by the powers of the brane width\footnote{In Ref.
\cite {lowe} it was found that on a D-brane in
bosonic string perturbation theory the induced
extrinsic curvature terms
are of the same order as the Einstein-Hilbert term and both
are suppressed by powers of string coupling constant.
However, the nonperturbative effects, e.g.,
those due to the inverse brane width studied above,
are expected to give the dominant contribution to
the worldvolume Einstein-Hilbert term.
The terms with extrinsic curvature will in general be
induced on a brane as well. It is interesting to study the
effects of these terms.}.

Therefore, we conclude that in the low-energy theory
the operators which would violate universality are  suppressed
by powers of $\Delta$. When the brane width is
small enough, of the order of  $1/\mpl$,
these  effects are unobservable at low energies.

After this digression we return back to the case
of one localized scalar field and try to study it
as far as possible.

Having the interaction vertices derived let us look at
the loop diagrams where two external graviton lines
are attached to the loops in which the localized fields are running.
The vertices in these diagrams are described by (\ref {g}).
One can think of this interaction as
a purely 4D interaction of $T_{\mu\nu}^\sigma(x)$
with the effective  4D graviton $h_{\mu\nu}(x)$.
The loop diagrams with two external
$h_{\mu\nu}$  lines give  rise to the following
term in the effective action on a brane:
\beq
{\tilde h}_{\mu\nu}(p)~{\hat {\cal O}}^{\mu\nu\alpha\beta}(p,M)
{\tilde h}_{\alpha\beta}(-p)~,
\label{kin}
\eeq
where ${\hat {\cal O}}^{\mu\nu\alpha\beta}(p,M)$
is some function of the external momentum $p$
and the particle mass $M$ (in fact, the UV cutoff in this case is
$M\sim \Delta^{-1}$). In the one loop approximation
this function  can be expressed in terms of hypergeometric
functions (see e.g, \cite {capper}).
This leads to a nonlocal interactions in the
effective Lagrangian in the coordinate  space. The explicit form of the
nonlocal coordinate-space Lagrangian is hard to present.
However, at
the momenta $p\ll M$ we can perform  an expansion of the
form-factor $ {\hat {\cal O}}^{\mu\nu\alpha\beta} $
in powers of $p^2/M^2$. The leading term takes the form:
\beqn
M^2 ~{\tilde h}_{\mu\nu}(p)\,\left (
\eta_{\mu\alpha}\eta_{\nu\beta}p^2\,-\,\eta_{\mu\nu}\eta_{\alpha\beta}p^2
\,-\,\eta_{\mu\alpha}p_\nu p_\beta  \right. \nonumber \\
\left.
-\, \eta_{\nu\beta} p_\mu p_\alpha
\,+\,\eta_{\alpha \beta} p_\mu p_\nu\, +\,
\eta_{\mu\nu} p_\alpha p_\beta \right )\,
{\tilde h}_{\alpha\beta}(-p)~.
\label{kinloc}
\eeqn
This is a quadratic approximation to a
reparametrisation invariant graviton
kinetic term  in a 4D low-energy effective theory of gravity
(for treatment of general relativity
as an effective field theory, see Ref. \cite {donoghue}).

Let us now turn to the  $\{i j\}$ components.
These will induce the brane-kinetic and mass terms for the components
$G_{ij}$. The latter, look as scalars (``graviscalars'')
from the point of view of a braneworld observer.
However, the mass  terms (potentials) of these states
on the brane might not be protected by any symmetries and
acquire  the mass of the order of the UV cutoff of the
worldvolume theory. Such states cannot mediate forces
which would compete with gravity at observable distances.
In addition, even if they stay massless,
they should decouple from the 4D matter at short (observable)
distances in analogy with the phenomenon found in
Ref. \cite {ddgv}.

Finally, we turn to the $\{i\mu\}$ components.
From a 4D perspective these are graviphotons.
In the linearized approximation they are
derivatively coupled to conserved currents.
These terms cannot give rise to gravity competing forces
at observable distances and
are irrelevant for our considerations.

\section{A Scalar Field Model}

To summarize the results of the previous section we write down
the gravitational part of the action which includes the lowest
dimensional derivative terms in the bulk and those induced
on the brane:
\beqn
S=\m^{2+N}~\int d^{4}x~d^Ny~\sqrt{|G|}~{\cal R}(G)~+~
M_{\rm ind}^{2}~\int d^{4}x
~\sqrt{|g|}~R(g)~+~{\rm{other~ terms}}~.
\label{actionG}
\eeqn
Here, the induced 4D term should be understood as an expansion
in perturbations defined as follows:
\beq
g_{\mu\nu}(x)~ \equiv~\eta_{\mu\nu}~+~h_{\mu\nu}(x)~,
\label{gh}
\eeq
where $h_{\mu\nu}(x)$ is related to the bulk graviton $G$
via (\ref {gg})\footnote{Note that the induced metric on the
brane is actually $A^2(\Delta)\eta_{\mu\nu}$.
The constant factor $A^2(\Delta)$ rescales $M^2_{\rm ind}$;
it is dropped for simplicity above. It will be taken into account
in  section 7.}. The first term in (\ref {actionG})
is the bulk kinetic term for a graviton and the second term
is a brane-induced kinetic term for the very same graviton.

In the previous section we found that in general the induced
terms are nonlocal. The term which is given in (\ref {actionG})
is just a first term in the expansion of that nonlocal expression
in powers of the brane width $\Delta$. However, as long as we
discriminate between  $h_{\mu\nu}(x)$ and $H_{\mu\nu}(x,0)$, the
induced term in (\ref {actionG}) itself contains the
effects of the same order, i.e., those which are proportional to
$\Delta$. This means that for any nonzero $\Delta$
the action (\ref {actionG}) is not complete.
The additional terms which vanish in the $\Delta \to 0$
limit are missing in  (\ref {actionG}).
Nevertheless, the latter action can be used to check the
limiting transition from the finite thickness brane to the delta function
brane since in the  limit  $\Delta \to 0$   the missing terms
do not contribute. Therefore, we will be able to check whether the
action (\ref {actionG}) gives rise to the results which in the limit
$\Delta \to 0$ turn smoothly to the ones obtained with a
delta function brane \cite{dgp,dg,dgs} answering some question raised in
Refs. \cite {kir,DR}.

Our goal is to study the interactions
between sources that  are located on the brane
and which exchange gravitons
described by the action (\ref {actionG}).
The value of the induced constant $M_{\rm ind}^{2}$
is crucial. In order to reproduce the correct 4D gravity on a brane
this has to equal to the 4D Planck mass,
$M_{\rm ind}^{2}=\mpl^2$ \cite{dgp,dg}. The question is how does one
get such a big scale on a  brane when the bulks scale $\m$ is
rather low. There are some known  possibilities for this
which we will mention briefly.
First, there exist certain branes,
even in 4D theories, the width of which is much smaller than
the fundamental length scale in the bulk (see discussions on this in Sect. 3
of Ref. \cite {dgkn1}.). In this case, the induced constant
will have the magnitude much bigger then the bulk scale $\m$.
More attractive possibility is to consider a
brane worldvolume particle theory
that has very high UV scale (such as the GUT scale).
States of the particle theory renormalize
the gravitational constant and make it of the order of
the UV cutoff.  In general, the induced scale is determined by a
two-point correlation function of the energy-momentum
tensor of the worldvolume particle theory which is localized on the
brane.  This was discussed in detail
in Refs. \cite {dgkn1} (see also \cite {dg}).
As a result, the 4D gravitational
Planck scale is a derived quantity.
Thus, the relation $M_{\rm ind}=\mpl$ is a
definition of the 4D Planck mass $\mpl$.

The next step is to calculate the two-point Green's functions
for the gravitational action (\ref {actionG}).
This calculation can be
straightforwardly performed for gravitons along the lines
of \cite {dgp}. However, to give more clear derivation
it is convenient to   make the following simplification.
In what follows we will
suppress the tensorial structure since the latter does  not change the
issue of singularities. Although, there will be separate
issues due to the  tensorial structure
(such as the ones associated with perturbative discontinuity
and nonperturbative continuity in $1/r_c$ advocated in
Ref. \cite {ddgv}). Since physics of higher codimension
cases is very similar to that of the 5D case,
we expect that these issues can be addressed  is a manner
similar to what was found for the 5D case in Ref. \cite {ddgv}
based on earlier work \cite {arkady}
(see also recent work \cite {lue}).
Having this said, in what follows we substitute the
graviton field by a scalar field \( \Phi  \)
and study the Green's functions for the latter.
This allows to determine the force-law. Hence,
the scalar $\Phi$ should be regarded as
a counterpart of a graviton.

Until section 7 we will work with a
brane which is placed at an orbifold fixed point and the
tension of which is tuned to zero by means of the
4D cosmological constant. This can be applicable, for instance,
to a non-BPS system of D-branes and Orientifold
planes placed on top of each other at certain fixed points.
The generalization to the nonzero tension branes
will be seen in section 7.
Before that, we should put
$\sqrt{G^{\rm cl}}=1,~A(y)=1$ and $\int f^2(y)d^Ny=1$.

The action for the scalar field which mimics
the graviton with the induced kinetic term on a brane in a space with
\( N \) extra dimensions is as follows:
\beqn
S ~= ~ \int \rm{d}^{4}x\rm{d}^{N}y~   \m^{2+N}~\left[
\partial _{A}\Phi (x,y)\partial ^{A}\Phi (x,y)~-~\frac{\epsilon  }
{\m^{2}}~
\Phi (x,y)
\left( \partial _{A}\partial ^{A}\right) ^{2}\Phi (x,y)\right]
 \nonumber \\
+~   M_{\rm{Pl}}^{2}~
\int \rm{d}^{4}x~\rm{d}^{N}y^{\prime}~ \rm{d}^{N}y^{\prime\prime}
~f^{2}(y^{\prime})~\partial _{\mu }\Phi (x,y^{\prime})
~f^{2}(y^{\prime\prime})~\partial ^{\mu }\Phi (x,y^{\prime\prime})~.
\label{action}
\eeqn
Note that we have
normalized the scalar field in such a way that it
is dimensionless.
The HD term, which is proportional to \( \epsilon   \),
is included for the regularization of the propagator (see below).
The parameter $\epsilon$ is an arbitrary ${\cal O}(1)$ constant.
Let us mention here that the results obtained in the
next sections are not bound to the  form of
the HD operator which we use. Any HD operator
which smoothes out the UV effects will reproduce
the same qualitative  results.

Below we discuss two separate cases.
First we approximate the  brane by delta-function and
show that singularities are removed by the HD terms.
In this case:
\beq
f^{2}(y)~=~\delta ^{(N)}(y)~.
\label{delta}
\eeq
As a next step  we consider a smooth nonzero thickness
brane and study the tree-level propagator.
For this we use the simplest ansatz:
\begin{equation}
\label{smoothbrane}
f^{2}(y)~=~\left\{ \begin{array}{cc}
\alpha ^{2},\,\, \, \rm{ for }~\left| y\right| < \Delta ,\\
0,\rm{ }\, \, \, \rm{for }~\left| y\right| > \Delta ~,
\end{array}\right.
\end{equation}
 where in the limit \( \Delta  \rightarrow 0,
\alpha ^{2}\rightarrow \infty  \)
and \( \alpha ^{2}\Delta ^{N}=const \).

\section{A Delta Function Brane}

In this section we restrict ourselves to the consideration of a
delta-function brane. The singularities will be removed by means
of the HD operators.
We adopt  the expression (\ref {delta}) for the localization function
and use the method of Ref. \cite{dg}. The equation  for the Green's
function for a source located on the brane takes the form
\beqn
\label{greendelta}
\left( \m^{2+N}~\left[\partial _{A}\partial ^{A}~+\frac{\epsilon  }
{\m^{2}}~\left( \partial _{A}\partial ^{A}\right)
^{2}\right]+~M_{\rm{Pl}}^{2}~\delta ^
{(N)}(y)~\partial _{\mu }\partial ^{\mu }\right)~
G(x,y;0,0) \nonumber \\
=\delta ^{(4)}(x)~\delta ^{(N)}(y)~.
\eeqn
To find a  solution of Eq. (\ref{greendelta}) let us turn to the
Fourier-transformed
quantities with respect to the worldvolume four-coordinates
\( x_{\mu } \):\[
G(x,y;0,0)\,=\,\int \frac{\rm{d}^{4}p}{\left( 2\pi \right)
^{4}}\,e^{ipx}\,\widetilde{G}\left( p,y\right)\, .\]
Then, in the Euclidean worldvolume space Eq. (\ref{greendelta})
takes the form
\beqn
\label{gpey}
\left( \m^{2+N}~\left[\left( p^{2}-\partial _{y}^{2}\right)~
+\frac{\epsilon  }{\m^{2}}
\left( p^{2}-\partial _{y}^{2}\right) ^{2} \right]+~
M_{\rm{Pl}}^{2}~\delta ^{(N)}(y)~p^{2}\right)
\widetilde{G}(p,y)~\nonumber \\
=~\delta ^{(N)}(y)~,
\eeqn
where \( p=\sqrt{p_{1}^{2}+p_{2}^{2}+p_{3}^{2}+p_{4}^{2}} \)
denotes the magnitude of the Euclidean momentum.

We are looking for the solution in the following form\[
\widetilde{G}(p ,y)~=~D(p ,y)~B(p ),\]
 where \( D(p ,y) \) is defined as follows\begin{equation}
\label{dpy}
\left( p ^{2}-\partial _{y}^{2}+\frac{\epsilon  }
{\m^{2}}\left( p ^{2}-\partial _{y}^{2}\right)
^{2}\right)\, D(p ,y)~=~\delta ^{(N)}(y)~.
\label{name}
\end{equation}
 \( B(p ) \) is some function which should  be determined.
Using the above expression one finds \cite {dg}:
\begin{equation}
\label{gpy}
\widetilde{G}(p ,y)~=~\frac{D(p ,y)}{
M_{\rm{Pl}}^{2}\,p ^{2}\,D(p ,0)\,+\,\m^{2+N} }\,,
\end{equation}
 where \( D(p ,y) \) is given by
\[
D(p ,y)\,=\,\int \frac{\rm{d}^{N}q}{\left( 2\pi \right) ^{N}}
\, \frac{{\rm exp}{(iqy)}}{p ^{2}\,+\,q^{2}\,+\,
\epsilon \,
\left( p ^{2}+q^{2}\right) ^{2}/\m^2 }\,.
\]
Note that for \( N\geq 2 \) the
\( {\epsilon  }/{\m^{2}} \)
term smoothes out the UV divergences in  the expression
for  \( D(p ,y) \).
Indeed, without this term the expression for \( D(p ,y) \)
blows up at \( y=0 \).
This divergence is not related to the presence of
the delta function brane  as it is clear from Eq. (\ref{dpy}).
Below we keep the HD term (in the \( N=1 \) case the
expression for \( D(p ,0) \) is finite even without the
HD term \cite{dgp,dgs,dg}.)

As an example of a theory with higher than one codimensions  we
consider the case of \( N=3 \). The corresponding calculation can be
done exactly.  The results for \( N \ge 2\) are similar
\footnote{Note that for \( N>3 \), in order to regularize
the function $D$, we should include higher
order operators beyond the one included already in (\ref {action}).}.

For \( N=3 \)  the function \( D(p ,y) \) is given by
\beqn
D(p ,y)\,=\,\frac{1}{4\pi\,|y| }\,
\left \{
{\rm exp}(-p \left| y\right| )
-\,{\rm exp} (-\left| y\right| {\tilde p})
\right \}~;~~~~~~~{\tilde p}~\equiv~
\sqrt{p ^{2}\,+\,\frac{\m^{2}} {\epsilon} }~.
\label{DD}
\eeqn
At \( y=0 \), we find
\beq
D(p ,0)~=~\frac{1}{4\pi }\,\left( \sqrt{p ^{2}\,+\,\frac{\m^{2}}
{\epsilon  }}\,-\,p \right)~\simeq~
\,\frac{1}{4\pi }\,\left( \frac{\m^{2}}{\epsilon  }\right)
^{\frac{1}{2}}\,\left( 1\,-\,p \,\left( \frac{\epsilon  }{\m^{2}}
\right) ^{\frac{1}{2}}\right)~,
\label{dd}
\eeq
which is finite. Note that, an
action with HD terms has a consistent interpretation
only if it is regarded  as an infinite series in derivatives.
Any truncation to a finite order can give rise to ghosts.
The HD term should be treated as
a $\left ({p ^{2}} /{\m^{2}}\right )$ correction in the expansion.
Hence, the above expression for  $D(p ,y) $  makes sense
only as an expansion in $\left ({p ^{2}} /{\m^{2}}\right )$
in which only the first correction is to be kept.
Combining this expression with  Eq. (\ref{gpy}), we get
\begin{eqnarray}
\widetilde{G}(p ,0)~\simeq ~ \frac{1}{\m^{4}\,\sqrt{\epsilon}\,
4\pi \left( 1\,+\,p \left( \frac{\epsilon  }{\m^{2}}\right)
^{\frac{1}{2}}\right)\, +\,M_{\rm{Pl}}^{2}\,p ^{2}}\,.\nonumber
\label{gpzero}
\end{eqnarray}
When the $ M_{\rm{Pl}}^{2}\,p ^{2}$ term in the
denominator dominates this gives the 4D Green's function.
The potential mediated by the scalar field \( \Phi  \)
on the 4D worldvolume
of the brane is determined as\[
V(r)=\int G(t,\overrightarrow{x},y=0;0,0,0)\rm{d}t,\]
 where \( r=\sqrt{x_{1}^{2}+x_{2}^{2}+x_{3}^{2}} \). Using Eq. (\ref{dd}),
we get
\begin{equation}
\label{potential}
V(r)\,=\,\int_0^{\infty}~ \frac{\rm{d}p}{\left( 2\pi \right) ^{2}}
\frac{2p\,\sin (pr)}{r}\,
\frac{1}{\m^{4}\,\sqrt{\epsilon  }\,
4\pi \left( 1\,+\,p \left( \frac{\epsilon  }{\m^{2}}\right)
^{\frac{1}{2}}\right)\, +\,M_{\rm{Pl}}^{2}\,p ^{2}}\,.
\end{equation}
This potential behaves as that of  a 4D theory
when  \( {\sqrt{\epsilon  }}/{\m} \ll r\ll \mpl/\m^2   \)
and behaves as the potential of a  \( (4+N) \)-dimensional theory
for bigger values of $r$. The
similar calculation for the
crossover distance can be done for other $N$'s.
The qualitative result is the same, i.e., the crossover distance
is approximated by:
\[
r^{(D\ge 6)}_{c}~\sim ~\frac{M_{\rm{Pl}}}{\m^{2}}.\]
 A simple estimate with \( M_{\rm{Pl}}\sim 10^{18}\rm{GeV} \)
and \( \m \sim 10^{-3}\rm{eV} \)
gives \( r_{c}\sim 10^{29}\rm{mm} \).
At the distance smaller than \( r_{c} \)
we will observe the four-dimensional
world. Thus, this model predicts
\textit{simultaneous} modification of gravity
at a distance of the order of \( 0.1 \) millimeter
and at around the Hubble distance!

Let us now study the case of two sources placed at different
positions \( y \) and \( y' \) in  extra dimensions.
The \( N=1 \) theory was studied in \cite{dgkn1}
(see also \cite {DR}).
In the present case the equation for the Green's
function in the momentum four-space takes the form:
\beqn
\label{gpeyyprime}
\left( \m^{5}\left[\left( p ^{2}\,-\,\partial _{y}^{2}\right)\,+\,
\frac{\epsilon  }
{\m^{2}}\,\left( p ^{2}\,-\,\partial _{y}^{2}\right) ^{2} \right]\,+
M_{\rm{Pl}}^{2}\,\delta ^{(3)}(y)\,p ^{2}\,\right)\,
\widetilde{G}(p ,y,y') \nonumber \\
=\,\delta ^{(3)}(y-y')\,.
\eeqn
The solution of Eq. (\ref{gpeyyprime}) can be written as follows:
\begin{eqnarray}
\widetilde{G}(p ,y,y')
~= ~\frac{1}{\m^{5}}\left( \frac{e^{-p \left| y-y'\right| }\,-\,
e^{-\left| y-y'\right| \sqrt{p ^{2}\,+\,\frac{\m^{2}}{\epsilon  }}}}
{4\pi \left| y-y'\right| }\right) \nonumber \\
-\,\frac{M_{\rm{Pl}}^{2}\,p ^{2}}{\m^{5}}\,
\left( \frac{e^{-p \left|
y\right| }-e^{-\left| y\right| \sqrt{p ^{2}+\frac{\m^{2}}{\epsilon  }}}}
{4\pi \left| y\right| }\right)\times
\left( \frac{e^{-p \left| y'\right| }-e^{-\left| y'\right|
\sqrt{p ^{2}+\frac{\m^{2}}{\epsilon  }}}}{4\pi \left| y'\right| }\right)
\label{sgyyprime} \\
 \times \frac{1}
{\m^{5}\,+\,{M_{\rm{Pl}}^{2}\,p ^{2}}\left( \sqrt{p ^{2}\,+\,\frac
{\m^{2}}{\epsilon  }}\,-\,p \right)/4\pi }~.
\nonumber
\end{eqnarray}
Using this expression one finds the potential:
\beqn
V(r,y,y')\, = \, \int G(t,\overrightarrow{x-x'},y,y')\,\rm{d}t\nonumber
\, =\,
 \int_0^{\infty} \frac{\rm{d}p}{\left( 2\pi \right)
^{2}}\frac{2p\,\sin (pr)}{r}~\widetilde{G}(p,y,y')\,,
\label{potentialyyprime}
\eeqn
where \( \widetilde{G}(p,y,y') \) is given by Eq. (\ref{sgyyprime}) .
All the expressions above should
be understood as an expansion in powers of
$\left (p^2/\m^2 \right )$. The expression (\ref {sgyyprime})
reveals that the bulk interactions near the brane
are strongly affected by the presence of the brane-induced
terms. This is similar
to what was observed in the $N=1$ case in Ref. \cite{dgkn1}.

From Eqs. (\ref{potentialyyprime},\ref{sgyyprime}),
it is clear that the potential between two sources placed
at different positions in  extra dimensions
does not give rise to any ultra-local interactions \cite{DR}.

\section{ Propagators on Thick Branes}

In this section we return to the consideration of branes
with a finite width. We will check whether
the results obtained for a delta function brane can
be reproduced from a thick brane by the limiting procedure
when $\Delta \to 0$.

The equation for the Green's function which is obtained
from (\ref {action}) takes the form:
\beq
{\hat {\cal  D}}~G(p,y)\,+\, \mpl^2\, p^2 \,f^2 (y)\,\int d^N y'\,
f^2(y')\,G(p,y')\,=\,g^2(y)\,,
\label{eqg}
\eeq
where $f^2$ and $g^2$ are two localized
functions with  different thicknesses
(we drop tilde sign over the momentum space Green's functions).
${\hat {\cal   D}}$ stands for  an operator in $N$ dimensions
which, as in the previous
section,  is approximated by the standard
d'Alambertian plus  one leading HD term:
\beq
{\hat {\cal  D}}\, =\,\m^{2+N} \,
\left \{  p^2\,-\,\nabla^2_N\,+\,\frac{\epsilon}{\m^2}\,
(p^2\,-\,\nabla^2_N)^2~\right \}\,.
\label{D}
\eeq
Before we turn to the calculation of the Green's function let us
make comments regarding the functions $f$ and $g$. The former
determines the localization of the breathing mode
on the brane worldvolume while the latter is a measure
of localization of a source under the consideration.
In general  these functions  might be different.
On the other hand, if we turn to the limit $\m\to 0$,
then  Eq. (\ref {eqg})  makes sense only if
$f\to g$ in that limit.
Thus $f$ and $g$ cannot be completely arbitrary functions,
there should exist some correlation between them.
This is a self-consistency requirement
which will play an important role below.

We look for a solution of Eq. (\ref {eqg}) in  the following form:
\beq
G(p,y)\,=\,G_0(p,y)~+~\chi(p,y)~,
\eeq
such that $G_0(p,y)$ obeys the equation
\beq
{\hat {\cal  D}}\,G_0(p,y)\,=\,g^2(y)\,,
\eeq
and, hence, can be written as follows:
\beq
G_0(p,y)~=~\int d^N y'~ g^2(y')D_*(p,y-y')~.
\eeq
Here $D_*=D/\m^{2+N}$,
denotes the Green's function for the  operator ${\hat {\cal D}}$
($D$ is defined in (\ref {name})).
Note also that in the above expression for $G_0$
we dropped the zero-mode solution of the operator
${\hat {\cal D}}$ which in this case is unphysical.

The equation for the function $\chi(p,y)$ takes the form:
\beq
{\hat {\cal  D}}\,
\chi(p,y)~+\mpl^2~ p^2~ [Y_1~+~Y_2]~f^2(y)~=~0~,
\eeq
where we introduced the following notations:
\beq
Y_1~\equiv~\int d^N y'f^2(y')~G_0 (p,y')~; ~~~ Y_2~\equiv~\int d^N y'f^2
(y')~\chi(p,y')~.
\label{ys}
\eeq
Disregarding the unphysical  solution
of the homogeneous equation we obtain:
\beq
\chi(p,y)\,=\,\mpl^2~ p^2 ~[Y_1~+~Y_2]~
\int d^N y'f^2(y')~D_*(p,y-y')\,.
\eeq
In order to determine $Y_2$ we integrate this equation
with the weight $f^2$.  Using the resulting
expression we derive the following equation for $\chi$:
\beq
\chi(p,y)~=~-~\frac{\mpl^2 \,p^2 ~[\int d^N y'f^2(y')G_0 (p,y')]~
~[\int d^N z~f^2(z)~D_*(p,y-z)] }
{1\,+\,\mpl^2\, p^2 \,\int d^N y ~d^N y'~ f^2(y)~D_*(p,y-y')
~f^2(y')}~.
\label{chi}
\eeq
Therefore,
the propagator in the general form reads as follows:
\beq
\label{prop}
G(p,y)~=~G_0 (p,y)~-~\frac{\mpl^2\, p^2 \,Y_1}{1\,+\,\mpl^2\,
p^2\, Y_3}\,Y(p,y)~,
\eeq
where we have used the notations introduced in (\ref {ys}) and
\begin{eqnarray}
Y_3 ~\equiv ~\int d^N y ~d^N y' ~f^2 (y)~D_*(p,y-y')~f^2(y')~,
\nonumber \\
Y(p,y) ~\equiv  ~ \int d^N y'~f^2(y')~D_*(p,y-y')~.
\end{eqnarray}

Below we will use these
general expressions for the Green's function  to derive
the properties of the potential on the brane.

Let us
study the $D=5$ case first.
We parametrize the functions $f$ and $g$
as follows:
\beq
f^2(y)~=~\left\{\begin{array}{ll}
\frac{1}{2\Delta} & \textrm{ for $|y|<\Delta$}~,\\
0 & \textrm{for $|y|>\Delta$}~; \end{array}
\right.\qquad
g^2(y)~=~\left\{\begin{array}{ll}
\frac{1}{2\alpha} & \textrm{ for $|y|<\alpha$}~,\\
0 & \textrm{for $|y|>\alpha$}~. \end{array} \right.
\eeq
First, for simplicity we assume that
$\alpha<\Delta$. Then,
inside the brane ($y<\Delta$) we find
\begin{eqnarray}
Y_1 & = & \frac{1}{2\Delta\,\m^3 \,p^2}
~\left(1-e^{-p\Delta}~\frac{\sinh(p\alpha)}{p\alpha}\right)~,
\nonumber \\
Y_3 & = & \frac{1}{2\Delta\,\m^3\, p^2}~
\left(1-e^{-p\Delta}~\frac{\sinh(p\Delta)}{p\Delta}\right)~,
\nonumber \\
Y(y) & = & \frac{1}{2\Delta\,\m^3\, p^2}~
\left(1-e^{-p\Delta}~\cosh(py)\right)~.
\end{eqnarray}
At $y=0$ the first order correction to the propagator reads
\beq
G(p,0)~\simeq ~\frac{1}{\mpl^2~p^2}~\left(1~+~\frac{p~\Delta}{6}~\right)~.
\label{6}
\eeq
As we discussed above (in a paragraph after Eq. (\ref {eqg})
the self-consistency of the
equations for the Green's function  (\ref {eqg})  requires that
$\Delta=\alpha+{\cal O}(\m/\mpl^2)$. This was used above
and the leading term was kept. Thus,
the correction due to the 4D propagator appears at the
scale $\Delta$. However, there
are other corrections due to the terms
of order ${\cal O}(\m/\mpl^2)$ which distinguish $\alpha$ and
$\Delta$. These terms can give rise to the
modification of the propagator at the momenta  above $\m$.
The latter modification is
expected to be in the theory anyway since
the higher dimensional operators are suppressed by $\m$.

As a  representative example of higher dimensions
we again concentrate on the codimension
three case, $N=3$. For the functions $f$ and $g$ we choose the following
interpolation:
\beq
f^2(y)~=~\left\{\begin{array}{ll}
\frac{3}{4\pi\Delta^3} & \textrm{ for $|\vec{y}|<\Delta$}~,\\
0 & \textrm{for $|\vec{y}|>\Delta$}~; \end{array} \right. \qquad
g^2(y)~=~\left\{\begin{array}{ll}
\frac{3}{4\pi\alpha^3} & \textrm{ for $|\vec{y}|<\alpha$}~,\\
0 & \textrm{for $|\vec{y}|>\alpha$}~. \end{array} \right.
\eeq
As before we set  $\alpha < \Delta$.
Somewhat  lengthy calculation
gives the following result for the Green's function
expanded in powers of the brane width $\Delta$:
\beq
G(p,0)~\simeq ~\frac{1}{\mpl^2~ p^2}~\left(1~+~\frac{39}{140}~p~\Delta
\right)~.
\label{39}
\eeq
As in the $N=1$  case we used the consistency relation
$\Delta=\alpha+{\cal O}(\m/\mpl^2)$.
The corrections to the 4D behavior arise at the scale
of order $1/\Delta$. However, this might be
changed if we were to take into account
a possible difference between
$\Delta$ and $\alpha$ which is of order  ${\cal O}(\m/\mpl^2)$.
The latter results in
the corrections to the propagator which
become significant at the scale $\m$.
Hence, the propagators
in $D=5$ and $D\ge 6$ are rather similar.

\section{The Crossover Scale}

In this section we will summarize the results on the crossover scale
$r_c$. This quantity is determined by the 4D Planck mass
$\mpl$ and the bulk Planck mass $\m$. As we discussed before,
in a 5D theory with a brane-induced Ricci term the expression
takes the form:
\beq
r^{({\rm D}=5)}_c~\sim~{\mpl^2\over \m^3}~.
\eeq
While in the case of higher dimensions with zero-tension branes
and the HD operators suppressed by powers of $1/\m$ we get:
\beq
r^{ ( {\rm D}\ge 6)}_c~\sim~{\mpl\over \m^2}~.
\label{rchi}
\eeq
Thus, in higher dimensions (i.e., for $D\ge 6$)
the value of $r_c$ is smaller.

Let us now discuss what are the
phenomenological bounds on $r_c$? The bounds could come from
different sources. Let us start with the upper bound.
There is a lower bound on the value of $\m$
which comes from the data on sub-millimeter gravity
measurements \cite {adel} and the accelerator, astrophysical and
cosmological data, that is
$\m\gsim 10^{-3}$ eV \cite {dgkn1}.
Hence, the upper bounds on $r_c$ are:
\beq
r^{({\rm D}=5)}_c \lsim 10^{59}
~{\rm mm}~; ~~~~~r^{({\rm D}\ge 6)}_c~\lsim
10^{29} ~{\rm mm}~.
\eeq
As we discussed in the introduction,
the latter number coincides with the present day horizon size.
Therefore, for $D\ge 6$ the model with
$\m \sim  10^{-3}$ eV predicts simultaneous
modification of laws of gravity at short distances
around $0.1$ mm and at large distances
around the Hubble scale.

Let us now turn to the lower bound on the crossover scale.
This bound can come from the measurements of the Newton
force at macroscopic distances, as well as from cosmological
considerations.

The distance at which Newton's law is known to hold
exceeds somewhat the solar system size. Beyond this  distance scale,
because of the presence of dark matter, the Newton
law can be modified  if the simultaneous changes are  made in
the amount and distribution of dark matter so that these changes
render a theory  consistent with the data on the
large scale structures. Therefore, from these arguments:
\beq
r^{\rm exp}_c ~\gg~ 10^{15}~{\rm cm}~.
\label{rexp}
\eeq
However, the cosmological consideration can impose a
stringer bound. Indeed, if the cosmological evolution changes at
distances of order $r_c$, then we want $r_c$ to be
of the order of the Hubble size at least.
The  cosmological solution of the  5D theory
which changes the regime  as the Hubble
parameter becomes of order $r_c$
was found by Deffayet in Ref. \cite {cedric}.
Therefore, we have to impose $r^{\rm exp}_c\gsim 10^{29}$ mm.
In this case, we  get $10^{-3}~{\rm eV}\lsim
\m \lsim  10^{7}~ {\rm eV}$ in the 5D theory. However,
in the $D\ge 6$ case we obtain a very stringent
restriction, $\m \sim 10^{-3}$ eV.
Thus, the fundamental scale of gravity in this model is bound to be
$\m \sim 10^{-3}$ eV.

Before we turn to the next section we would like to
make some comments.
Recently new interesting branches of cosmological solutions were found
by Dick \cite {dick,dick1}, and by Cordero and Vilenkin
\cite {vilenkin}. The creation of the  ``stealth'' branes were
discussed in \cite {vilenkin}.
These solutions, from the perspective of a 4D braneworld observer,
coincide with the well known solutions of pure 4D theory.
On this branch,  there is no difference between the
time evolution of the universe in just a
purely 4D theory from that in a  higher dimensional theory
discussed in the present paper. These solutions are found as follows
\cite {dick,dick1,vilenkin}. Consider the Einstein equations with
the induced terms included:
\beqn
\m^{2+N}\,\left ({\cal R}_{AB}\,-\,{1\over 2}G_{AB}\,{\cal R}\right )
\, +\, M_{\rm ind}^2\,\delta^\mu_A \delta^\nu_B \, \delta^{(N)}(y)\,
\left (R_{\mu\nu}\,-\,{1\over 2}g_{\mu\nu}\,R \right )\,  \nonumber \\
=\,\delta^\mu_A \delta^\nu_B \,    T^{\rm Brane}_{\mu\nu}~\delta^{(N)} (y)~.
\label{eeq}
\eeqn
There can exist a solution which is flat in the bulk but
satisfies the equation on the brane, i.e.:
\beqn
{\cal R}_{AB}~=~0~,\nonumber \\
\label{4dsol}
M_{\rm ind}^2
\left (R_{\mu\nu}\,-\,{1\over 2}g_{\mu\nu}\,R \right )\, =\, T^{\rm
Brane}_{\mu\nu}~.
\label{4dsolut}
\eeqn
The solution of the second  equation in this system is
exactly the one of a pure 4D theory. It
is also a solution to the whole 5D system provided that
the metric is flat in higher dimensions
(see Refs. \cite {dick,dick1}, \cite {vilenkin}).

If this branch is realized,
then there is no bound on $r_c$ arising from cosmology.
The only restriction on $r_c$ comes \cite {dick1} from
the solar system measurements of Newton's law (\ref {rexp}).
This can be translated into the following
upper bounds on $\m$:
\beqn
10^{-3}~{\rm eV}\lsim \m \lsim 10^{12}~{\rm eV}~, ~~~{\rm for}~~ D=5~;
\nonumber \\
10^{-3}~{\rm eV}\lsim \m \lsim 10^{4}~{\rm eV}~,~~~{\rm for}~~ D\ge 6~.
\label{constr}
\eeqn
As before, there is a bigger range for $\m$ in the 5D theory.

\section{Comments on Nonzero Tension Branes}

In the previous sections we dealt with branes
(or system of branes) which were placed at orbifold fixed points
and for which the tension effects were removed by fine tuning.
Here we would like to study the influence
of a nonzero brane tension on the expression
for $r_c$.
We start with an infinite $(4+N)$ dimensional space.
Consider a certain
brane in this space which has an intrinsic tension $T$.
The brane distorts the ambient space around it.
For a BPS D3-brane the solutions is regular outside of the core
and has well defined horizon. However, for non-BPS branes
which are phenomenologically relevant, the solutions
give rise to singularities. The singularity at the core
is an  artifact  of the delta-function approximation is
expected to be removed if the core of the brane is smoothed
\cite {Gregory} (see also \cite {Emparan}).
We will adopt below this philosophy.

For a brane world which preserves 4D Lorentz invariance
the interval takes the form:
\beq
ds^2~=~A^2(y)~\eta_{\mu\nu}dx^\mu dx^\nu~-~B^2(y)~
dy^2~-C^2(y)\,y^2~d\Omega^2_{N-1}~,
\label{interval}
\eeq
where the functions $A$, $B$ and $C$ depend on the  brane
tension and $\m$. For certain cases
$A,B$ and $C$ are known exactly ( see, e.g., Refs.
\cite {Duff,Gregory,Oz}).
In the $N=2$ case
a positive-tension brane creates a deficit angle and produces no other
nontrivial warp-factors like $A,B$ and $C$.
The effects of this on the model with brane-induced Einstein-Hilbert
term were studied in Ref. \cite {zura2} where it was concluded that
the qualitative behavior found in \cite {dg} (and outlined above)
does not change.

Below we will concentrate on the case $N\ge 3$.
We use the solutions  for non-BPS branes
obtained in Ref. \cite {Gregory}:
\beq
A=f^{-{1\over 4} \sqrt{N-1\over N+2} }~,~~B= f^{-{1\over N-2}
\left ( {N-3\over 2} - \sqrt{N-1\over N+2} \right )}~,~~
C=f^{{1\over N-2}\left ({1\over 2}+ \sqrt{N-1\over N+2} \right )}~,
\label{background}
\eeq
where
\beq
f \equiv 1~+~\left (y_g\over y  \right )^{N-2}~.
\eeq
Here the gravitational radius of a brane in the
transverse direction $y_g$ is determined by the brane
tension $T$ and can be written as follows
\beq
y_g~=~\left ( {T\over \m^4}  \right )^{1\over N-2}~{1\over \m}~.
\label{rg}
\eeq
In order to determine the
properties of the interactions on the brane
one needs to calculate a graviton
propagator on the background  (\ref {background}).
The exact calculation is cumbersome
(for certain calculations
along these directions see Ref. \cite {Emparan}).
However, one can do a qualitative analysis of the
equation for perturbations. The analysis is similar to that
performed in Ref. \cite {dg} and will not be repeated here.
We just mention the basic  steps.

The solutions for $A, B$ and $C$ in (\ref {background}) are singular
near $y=0$. However, in this domain the classical theory
breaks down and the expressions  (\ref {background})
should be smoothed out.
In the previous section we accounted for this UV
regularization by introducing higher dimensional operators.
This is perhaps the most consistent and reliable
way of dealing with the singularities.
However, when a nonzero brane tension is taken into account,
the higher dimensional operators complicate
the equations even further so that they
are not treatable analytically. Under these circumstances, it
becomes more reasonable to introduce other UV regularization,
e.g., one could use  the finite width of the brane core $\Delta$.
Then, we can estimates the
critical momentum/distance  for which the 4D induced term becomes
dominant over the bulk terms in the background (\ref {background}).
Note that this approach has a rather limited applicability since
it is justified only when  $T\sim \m^4$.
This consideration leads to the following
region where the effects  of the 4D induced terms dominate:
$r~\lsim ~r^{(T\sim \m^4)}_c~\sim ~{\mpl / \sqrt{T ~+~\Delta^{N-2}~
\m^{N+2}}}$.
Here the 4D Planck mass is determined by the product of the
induced scale and the value of the function $A$ taken in the core
$\mpl^2 =M_{\rm ind}^2A^2({\Delta})$. Although, the value of
$A^2(\Delta)$ is not known, one might expect that it
is not far away from the value of this function at
the point  where the gravity approximation is reliable.
Thus, it adds/subtracts an order of magnitude from $M_{\rm ind}$.
Note also that since for a phenomenologically viable model
$\m \sim 10^{-3} ~{\rm eV}$, then the expression for
$r^{(T\sim \m^4)}_c$ is applicable only for
$T\le (10^{-3} ~{\rm eV})^4$.

Let us now turn to the constructions in which
our brane is represented by a stuck of branes and anti-branes
(or orientifold planes)
which can be placed at some orbifold fixed points.
In this case one can study consistently the limit $T\to 0$.
The resulting expression for $r^{(T\sim \m^4)}_c$
coincides with  that obtained in Ref. \cite {dg}.
The crossover scale depends in general on the
number of dimensions.  On the other hand,
in the previous sections we studied a theory with high derivatives
and obtained the crossover scale which was independent of
$N$ (\ref {rchi}). The question is how these two results are
reconciled.
The reconciliation is based on the following arguments.
When  we  take  into account
the higher dimensional operators which are suppressed by $\m$,
this means that UV gravitational resolution of the brane
width is at the scale $\Delta~\sim~1/\m$. Substituting the
latter expression in the expression for $r^{(T\sim \m^4)}_c$
with $T=0$ we recover  Eq. (\ref {rchi}). Thus, the phenomenological
constraints on the bulk gravitational
scale are the same as in the previous section.

Finally, the following scenario is also possible:
suppose the bulk gravitational scale is high,
let us say in a TeV region  or so.
Let us call this scale  $M_{\rm Bulk} \gsim {\rm TeV}$.
It is conceivable that some scalar field dynamics suppresses the
coefficient in front of the Ricci scalar in the bulk. Then,
$\m^{4+N}=f(\langle \Phi \rangle)M^{4+N}_{\rm Bulk}$,
where $f(\langle \Phi \rangle)$ is a suppression factor
which depends on the vacuum expectation value  of a scalar.
On the other hand, the higher derivative terms in the
bulk are suppressed by  the scale $M_{\rm Bulk}\gg \m$.
Let us consider a zero-tension brane at the
orbifold fixed point. Then, in the expression for
$r^{(T\sim \m^4)}_c$  we should
take $\Delta \sim 1/M_{\rm Bulk}$.
Using the hierarchy $M_{\rm Bulk}\gg \m$ we can obtain
an acceptable values of $r_c$ even for $\m \sim {\rm TeV}$.
For instance if $M_{\rm Bulk} \sim  \mpl$ we get
the value of $r_c$ which is consistent with (\ref {rexp})
for $N\ge 4$.

\section{Conclusions}

Summarizing, we discussed a brane-world model with
two or more infinite-volume extra dimensions.
This model  predicts the simultaneous modification of the
law of gravity at around 0.1 mm and at the Hubble scale.
Therefore, the fundamental scale of gravity is
bound to be $10^{-3}$ eV in this model.
These predictions can be tested by table-top
gravitational measurements
and by astrophysical experiments.

The UV singularities which arise in this model can be treated
consistently. We used both, the regularization by a finite brane
width,  and by the higher-dimensional
operators which are suppressed by the fundamental scale of gravity.
We calculated the corresponding two-point
Green's functions in the regularized theory
and  showed that there is an intimate relation
between the UV and IR scales where  gravity
changes its regime.

\section{Acknowledgments}

We are grateful to S. Dimopoulos,
A. Gruzinov, D. Kabat, F. Nitti, M. Perelstein, R. Rattazzi,
M. Shifman, A. Vainshtein, M. Voloshin and M. Wise
for useful discussions.
GD and GG thank {\it The Aspen Center for Physics},
and GG thanks CERN Theory Division for the hospitality
where parts of this work were done.
The work of GD was supported in part by David and Lucille Packard
Foundation Fellowship for Science and Engineering, by Alfred P. Sloan
foundation fellowship and by NSF grant PHY-0070787.
The work of GG and XH is supported by  DOE Grant DE-FG02-94ER408.


\newpage


\begin{thebibliography}{10}

\bibitem{adel}
C.~D.~Hoyle, U.~Schmidt, B.~R.~Heckel,
E.~G.~Adelberger, J.~H.~Gundlach, D.~J.~Kapner and H.~E.~Swanson,
%``Sub-millimeter tests of the gravitational inverse-square law:
%A search  for 'large' extra dimensions,''
Phys.\ Rev.\ Lett.\  {\bf 86}, 1418 (2001)
[arXiv:hep-ph/0011014];\\
%%CITATION = HEP-PH 0011014;%%
J.C. Price, in {\it Proceedings of
International Symposium on Experimental Gravitational Physics}, ed. P.
Michelson, Guangzhou, China (World Scientific, Singapore,1988);\\
J. Long, ``Laboratory Search for Extra-Dimensional Effects in
Sub-Millimeter
Regime'' Talk given at the International Conference on Physics Beyond Four
Dimensions, ICTP, Trieste, Italy; July 3-6, (2000);\\
A. Kapitulnik, ``Experimental Tests of Gravity Below 1mm''
Talk given at the International Conference on Physics Beyond Four
Dimensions, ICTP, Trieste, Italy; July 3-6, (2000)~.


\bibitem{cc} A.G. Riess et al., {\it Astroph. J} 116, 1009 (1998);\\
S. Perlmutter et al., ``Measurements of Omega
and Lambda from 42 High-Redshift Supernovae", [astro-ph/9812133];\\
A.G. Riess, Talk Given at The Symposium ``{\it The Dark Universe:
Matter, Energy, and Gravity}'' Baltimore, April 2-5, (2001).


\bibitem{dgp}
G.~R.~Dvali, G.~Gabadadze and M.~Porrati,
%``4D gravity on a brane in 5D Minkowski space,''
Phys.\ Lett.\ B {\bf 485}, 208 (2000)
[arXiv:hep-th/0005016].
%%CITATION = HEP-TH 0005016;%%



\bibitem{dgkn1}
G.~R.~Dvali, G.~Gabadadze, M.~Kolanovic and F.~Nitti,
%``Scales of gravity,''
arXiv:hep-th/0106058;\\
%%CITATION = HEP-TH 0106058;%%
Phys.\ Rev.\ D {\bf 64}, 084004 (2001)
[arXiv:hep-ph/0102216].
%%CITATION = HEP-PH 0102216;%%




\bibitem{dg}
G.~R.~Dvali and G.~Gabadadze,
%``Gravity on a brane in infinite-volume extra space,''
Phys.\ Rev.\ D {\bf 63}, 065007 (2001)
[arXiv:hep-th/0008054].
%%CITATION = HEP-TH 0008054;%%




\bibitem{cedric}
C.~Deffayet,
%``Cosmology on a brane in Minkowski bulk,''
Phys.\ Lett.\ B {\bf 502}, 199 (2001)
[arXiv:hep-th/0010186].
%%CITATION = HEP-TH 0010186;%%


\bibitem{ddg}
C.~Deffayet, G.~R.~Dvali and G.~Gabadadze,
%``Accelerated universe from gravity leaking to extra dimensions,''
arXiv:astro-ph/0105068;
%%CITATION = ASTRO-PH 0105068;%%
%C.~Deffayet, G.~R.~Dvali and G.~Gabadadze,
%``Comments on ``A Supernova Brane Scan'',''
arXiv:astro-ph/0106449.
%%CITATION = ASTRO-PH 0106449;%%





\bibitem{dick} R.~Dick,
%``Brane worlds,''
Class.\ Quant.\ Grav.\  {\bf 18}, R1 (2001)
[arXiv:hep-th/0105320].
%%CITATION = HEP-TH 0105320;%%

\bibitem{dick1}
R.~Dick,
%``Standard cosmology in the DGP brane model,''
arXiv:hep-th/0110162.
%%CITATION = HEP-TH 0110162;%%

\bibitem{vilenkin}
R.~Cordero and A.~Vilenkin,
%``Stealth branes,''
arXiv:hep-th/0107175.
%%CITATION = HEP-TH 0107175;%%


\bibitem{kofinas}
G.~Kofinas,
%``General brane cosmology with (4)R  term in (A)dS(5) or Minkowski bulk,''
JHEP {\bf 0108}, 034 (2001)
[arXiv:hep-th/0108013].
%%CITATION = HEP-TH 0108013;%%


\bibitem{ddgv}
C.~Deffayet, G.~R.~Dvali, G.~Gabadadze and A.~I.~Vainshtein,
%``Nonperturbative continuity in graviton
%mass versus perturbative  discontinuity,''
arXiv:hep-th/0106001.
%%CITATION = HEP-TH 0106001;%%


\bibitem{ian}
I.~Giannakis and H.~c.~Ren,
%``Linearized analysis of the Dvali-Gabadadze-Porrati brane model,''
arXiv:hep-th/0111127.
%%CITATION = HEP-TH 0111127;%%

\bibitem{lue}
A.~Lue,
%``Cosmic strings in a brane world theory with metastable gravitons,''
arXiv:hep-th/0111168.
%%CITATION = HEP-TH 0111168;%%



\bibitem{dgs}
G.~R.~Dvali, G.~Gabadadze and M.~A.~Shifman,
%``(Quasi)localized gauge field on a brane:
%Dissipating cosmic radiation  to extra dimensions?,''
Phys.\ Lett.\ B {\bf 497}, 271 (2001)
[arXiv:hep-th/0010071].
%%CITATION = HEP-TH 0010071;%%


\bibitem{akhmedov}
E.~K.~Akhmedov,
%``Dynamical localization of gauge fields on a brane,''
Phys.\ Lett.\ B {\bf 521}, 79 (2001)
[arXiv:hep-th/0107223].
%%CITATION = HEP-TH 0107223;%%


\bibitem{Kogan}
I.~I.~Kogan, S.~Mouslopoulos, A.~Papazoglou, G.~G.~Ross and J.~Santiago,
%``A three three-brane universe:
%New phenomenology for the new  millennium?,''
Nucl.\ Phys.\ B {\bf 584}, 313 (2000)
[arXiv:hep-ph/9912552].
%%CITATION = HEP-PH 9912552;%%

\bibitem{GRS}
R.~Gregory, V.~A.~Rubakov and S.~M.~Sibiryakov,
%``Opening up extra dimensions at ultra-large scales,''
Phys.\ Rev.\ Lett.\  {\bf 84}, 5928 (2000)
[arXiv:hep-th/0002072].
%%CITATION = HEP-TH 0002072;%%


\bibitem{sundrum}
R.~Sundrum,
%``Towards an effective particle-string
%resolution of the cosmological  constant problem,''
JHEP {\bf 9907}, 001 (1999)
[hep-ph/9708329].
%%CITATION = HEP-PH 9708329;%%

\bibitem{dgpcc}
G.~R.~Dvali, G.~Gabadadze and M.~Porrati,
%``Metastable gravitons and infinite volume extra dimensions,''
Phys.\ Lett.\ B {\bf 484}, 112 (2000)
[arXiv:hep-th/0002190];
%%CITATION = HEP-TH 0002190;%%
%G.~R.~Dvali, G.~Gabadadze and M.~Porrati,
%``A comment on brane bending and ghosts
%in theories with infinite extra  dimensions,''
Phys.\ Lett.\ B {\bf 484}, 129 (2000)
[arXiv:hep-th/0003054].
%%CITATION = HEP-TH 0003054;%%

\bibitem{witten}
E.~Witten,
``The cosmological constant from the viewpoint of string theory,''
{\it Lecture given at 4th International Symposium on Sources and
Detection of Dark Matter in the Universe};
Marina del Rey, California, 23-25 Feb 2000.
[arXiv:hep-ph/0002297].
%%CITATION = HEP-PH 0002297;%%

\bibitem{kir}
E.~Kiritsis, N.~Tetradis and T.~N.~Tomaras,
%``Thick branes and 4D gravity,''
JHEP {\bf 0108}, 012 (2001)
[arXiv:hep-th/0106050].
%%CITATION = HEP-TH 0106050;%%


\bibitem{lowe}
S.~Corley, D.~A.~Lowe and S.~Ramgoolam,
%``Einstein-Hilbert action on the brane for the bulk graviton,''
JHEP {\bf 0107}, 030 (2001)
[arXiv:hep-th/0106067].
%%CITATION = HEP-TH 0106067;%%


\bibitem{zura}
Z.~Kakushadze,
%``Orientiworld,''
JHEP {\bf 0110}, 031 (2001)
[arXiv:hep-th/0109054].
%%CITATION = HEP-TH 0109054;%%


\bibitem{wagner}
M.~Carena, A.~Delgado, J.~Lykken, S.~Pokorski, M.~Quiros and C.~E.~Wagner,
%``Brane effects on extra dimensional scenarios: A tale of two gravitons,''
Nucl.\ Phys.\ B {\bf 609}, 499 (2001)
[arXiv:hep-ph/0102172].
%%CITATION = HEP-PH 0102172;%%


\bibitem{DR}
S.~L.~Dubovsky and V.~A.~Rubakov,
%``On models of gauge field localization on a brane,''
Int.\ J.\ Mod.\ Phys.\ A {\bf 16}, 4331 (2001)
[arXiv:hep-th/0105243].
%%CITATION = HEP-TH 0105243;%%


\bibitem{capper}
D.M. Capper, Nuovo Cim. {\bf A25} 29 (1975).

\bibitem{donoghue}
J.~F.~Donoghue,
%``General Relativity As An Effective Field Theory:
%The Leading Quantum Corrections,''
Phys.\ Rev.\ D {\bf 50}, 3874 (1994)
[arXiv:gr-qc/9405057].\\
%%CITATION = GR-QC 9405057;%%
J.~F.~Donoghue,
``Introduction to the Effective Field Theory Description of Gravity,''
{\it Talk given at Advanced School on Effective Theories},
Almunecar, Spain, 25 Jun - 1 Jul 1995. [arXiv:gr-qc/9512024].
%%CITATION = GR-QC 9512024;%%



\bibitem{arkady} A.I. Vainshtein, Phys. Lett.
{\bf 39B}, 393 (1972)~.

\bibitem{Gregory}
R.~Gregory,
%``Cosmic p-Branes,''
Nucl.\ Phys.\ B {\bf 467}, 159 (1996)
[arXiv:hep-th/9510202].
%%CITATION = HEP-TH 9510202;%%

\bibitem{Emparan}
C.~Charmousis, R.~Emparan and R.~Gregory,
%``Self-gravity of brane worlds: A new hierarchy twist,''
JHEP {\bf 0105}, 026 (2001)
[arXiv:hep-th/0101198].
%%CITATION = HEP-TH 0101198;%%

\bibitem{Duff} M.~J.~Duff, R.~R.~Khuri and J.~X.~Lu,
%``String solitons,''
Phys.\ Rept.\  {\bf 259}, 213 (1995)
[arXiv:hep-th/9412184].
%%CITATION = HEP-TH 9412184;%%

\bibitem{Oz} O.~Aharony, S.~S.~Gubser, J.~Maldacena, H.~Ooguri and Y.~Oz,
%``Large N field theories, string theory and gravity,''
Phys.\ Rept.\  {\bf 323}, 183 (2000)
[arXiv:hep-th/9905111].
%%CITATION = HEP-TH 9905111;%%

\bibitem{zura2}
O.~Corradini, A.~Iglesias, Z.~Kakushadze and P.~Langfelder,
%``Gravity on a 3-brane in 6D bulk,''
Phys.\ Lett.\ B {\bf 521}, 96 (2001)
[arXiv:hep-th/0108055].
%%CITATION = HEP-TH 0108055;%%


\end{thebibliography}
\end{document}